\documentclass[a4paper,12pt]{article}
\usepackage[centertags]{amsmath}
\usepackage{amssymb}

\usepackage{graphicx}
\usepackage{epsfig}
\usepackage{ulem}
\usepackage[english]{babel}
\usepackage{array}
\usepackage{amsthm}
\usepackage{latexsym}

\usepackage{yfonts}
\usepackage{mathrsfs}
\usepackage[mathcal]{euscript}

\pdfoutput=1
\usepackage{epsfig}
 \usepackage{jheppubm}
\usepackage{mathdots}
\usepackage{MnSymbol}
\usepackage{multirow}

\definecolor{darkblue}{rgb}{0., 0, 1}

\definecolor{dgreen}{rgb}{0.,0.6,0.}

\definecolor{dgreen}{rgb}{0.,0.6,0.}

\newcommand{\nn}{\nonumber}
\newcommand{\be}{\begin{equation}}
\newcommand{\ee}{\end{equation}}
\newcommand{\bea}{\begin{eqnarray}}
\newcommand{\eea}{\end{eqnarray}}

\definecolor{dgreen}{rgb}{0.,0.6,0.}

\title{Cosmological Constant and \\
Maximum of Entropy for   de Sitter Space\\  
} 

\author{Igor Volovich}

 \affiliation{Steklov Mathematical Institute, Russian Academy of Sciences}
 \emailAdd{volovich@mi-ras.ru}

\abstract{
There are at least two cosmological constants calling for explanation.
The first one describes the quasi-de Sitter inflation in the early universe, and the second describes the current acceleration of the universe associated with dark energy.  
An approach to the computation of the inflationary  cosmological constant in the early universe is proposed. The tunneling and no-boundary proposals suggest that  the ground state of the  early universe is the de Sitter space, which is  the maximal  symmetric  spacetime.
In this paper it is argued that the radius of the de Sitter space, i.e. the cosmological constant, can be computed using the principle of maximum entropy.
The universe emerges  from ``nothing" that corresponds to a minimum of entropy.  The entropy reaches its maximal value for the 4-dimensional de Sitter space with the inflationary cosmological constant $\Lambda=3\pi\,\exp\{-\psi(3/2)\}$, where $\psi$ is the digamma function and 
 $\Lambda\approx 9.087 $  in Planck units. 
 
}
\begin{document}

\maketitle
\section{Introduction}
Einstein introduced the cosmological constant \cite{Einstein} assuming that there is only one cosmological constant, since there is only one universe. However now it is assumed that there are at least two cosmological constants, the inflationary cosmological constant describing the quasi-de Sitter phase in the early universe and the  second  cosmological constant
describing the current acceleration of the universe 
associated with dark energy.
\\

A possible connection between a cosmological constant and elementary particle physics was considered by Zeldovich \cite{Zeldovich:1968ehl}. In particular he suggested, based on Sakharov's previous work \cite{Sakharov:1967pk}, a possible connection between the vacuum fluctuations of quantum matter and the cosmological constant.  The inflationary cosmology is considered in \cite{Starobinsky80,Guth81,Linde82}.
\\

 In the present paper an approach to the computation of the inflationary cosmological constant in the early universe is proposed. The tunneling \cite{Vil} and no-boundary \cite{HH}  proposals suggest that  the ground state of the universe is  de Sitter space.  In this paper it is argued that the radius of the de Sitter space, i.e. the cosmological constant in the original action, can  be computed if one   uses the  principle of  maximum entropy and also  consider the spacetime dimension as a continuous parameter of the theory. Note that the consideration of the spacetime dimension as a continuous parameter is widely used as a regularization in quantum field theory \cite{tHooft:1973wag} and also in theories of phase transitions \cite{Wilson:1973jj}.
\\

In   Sect.\ref{sect:CC}, it will be shown that the universe, arising from ``nothing" with minimal entropy, reaches, for the 4-dimensional de Sitter space, the maximum entropy value for the cosmological constant given by
  \be
  \label{Lambda9}\Lambda=
   3\pi\,\exp\{-\psi(3/2)\},\ee
   where $\psi$ is the digamma function,
 $\Lambda\approx 9.087 $  in Planck units.  For $0\leq D\leq4$, entropy at $\Lambda$ given by \eqref{Lambda9} is a monotonically increasing function of the number of dimensions, which can be said to play the role of time.
\\

The proposal that the area of the cosmological horizon should be interpreted as an entropy was made in \cite{GH}. It is proposed that the entropy of the empty de Sitter space maximises the de Sitter entropy  \cite{Mae,Bou}. There exists a maximum entropy state on de Sitter space. Time dependence of entropy of de Sitter space is considered in \cite{Frolov:2002va,Chen:2012bi}.
An algebra of observables of type $II_1$ describing the maximum entropy state in the de Sitter space is given in \cite{CLPW}.
The correspondence between  de Sitter space and Bose gases is considered in \cite{AV}.
Note that the expression for the entropy even for positive $D$  includes the area of the unit sphere $S^{D-2}$,
so the negative  dimensions are involved, see Sect.\ref{CND} and \cite{AV}.
\\

The paper is organized as follows. 
In Sect.\ref{CND} remarks on  complex and negative dimensions are made. Cosmological constant in early universe is computed in Sect.\ref{sect:CC}.
In conclusion initial data for inflation and the Planck era are briefly discussed.

 \section{Cosmological Constant}
 \label{sect:CC}
 In this Section the inflationary cosmological constant $\Lambda$ for the early universe is computed.
 \subsection{Preliminary}
 The de Sitter metric has the form
 \be
  ds^2=-dt^2 +e^{2Ht}\left(dy_1^2+...+dy_{D-1}^2\right),\ee
 where the Hubble parameter $H$ is related with the cosmological constant as
 \be
 \Lambda=\frac12 (D-1)(D-2) H^2.\ee
 The metric on the static patch of de Sitter space is
\be\nn
 ds^2=-\left(1-\frac{r^2}{\ell^2}\right)dt^2+\left(1-\frac{r^2}{\ell^2}\right)^{-1}dr^2+r^2 d\omega_{D-2}^2.
\ee
where the de Sitter radius $\ell$ is $\ell=1/H$.
 The Gibbons-Hawking \cite{Gibbons:1977mu} temperature and the corresponding entropy are
 \bea
 T&=&\frac{1}{2\pi\ell},\\
S&=&  S(\ell,D)= \frac{\ell^{D-2}}{4}\Omega _{D-2},
 \eea
here $\ell$ is the radius of $D$-dimensional de Sitter space and  $\Omega _{D-2}$ 
    is the area of the sphere $S^{D-2}$ 
\be\Omega _{D-2}=\frac{2 \pi ^{(D-1)/2}}{\Gamma(\frac{D-1}{2})}.\ee

\subsection{Principle of maximum entropy} \label{sect:PME}

 The principle of maximum entropy states that the probability distribution which best represents the current state of knowledge about a system is the one with largest entropy, see  \cite{Jaynes:1957zza}. \\
 
 We propose to  extend the  principle of maximum entropy and include, as a possible implementation of a physical system, a consideration of the system in the spacetime of an arbitrary dimension.  We can then use the empirical knowledge that the observed dimension of spacetime is $D_0=4$ and try to find a realization of the modified maximum entropy principle that guarantees that the dimensionality of spacetime is $D_0=4$.  Application of the extend principle of maximum entropy to the de Sitter spacetime gives the following.
 \\
 
 We take the  value of entropy of D-dimensional de Sitter space with the radius $\ell$,  $S(\ell,D)$,  vary the spacetime dimension D and find  
 location of the maximum. It happens that the value of $D$ where the maximum is realized, $D_{max}$ depends on $\ell$
 \be \label{ell0}
 \max_{D}\,S(\ell,D)=S(\ell,D_{\max}(\ell)).
 \ee
 The idea is to find such $\ell=\ell_0$ that $D_{\max}(\ell_0)=4$.

 \subsection{Computation of $\Lambda$} 
The entropy of  de Sitter space is given by 
\be S(\ell,D)=\frac{1}{4 }\,\ell^{D-2} \Omega _{D-2}=\frac{1}{2 }\,\ell ^{D-2}\frac{\pi^{(D-1)/2}}{\Gamma((D-1)/2)}.
\ee
We consider $D$ as a real variable. 
The derivative of entropy with respect to $D$ is 
\bea
\frac{\partial  S(\ell,D)}{\partial D}=\frac{\pi ^{\frac{D-1}{2}} \ell^{D-2}}{4 \Gamma \left(\frac{D-1}{2}\right)} \left(\log (\pi \ell^2)-\psi
   \left(\frac{D-1}{2}\right)\right),\eea
where $\psi
   $ is the digamma function, see eq.\ref{psi}. 
\\

\begin{figure}[h]
    \centering
    \includegraphics[width=70mm]
{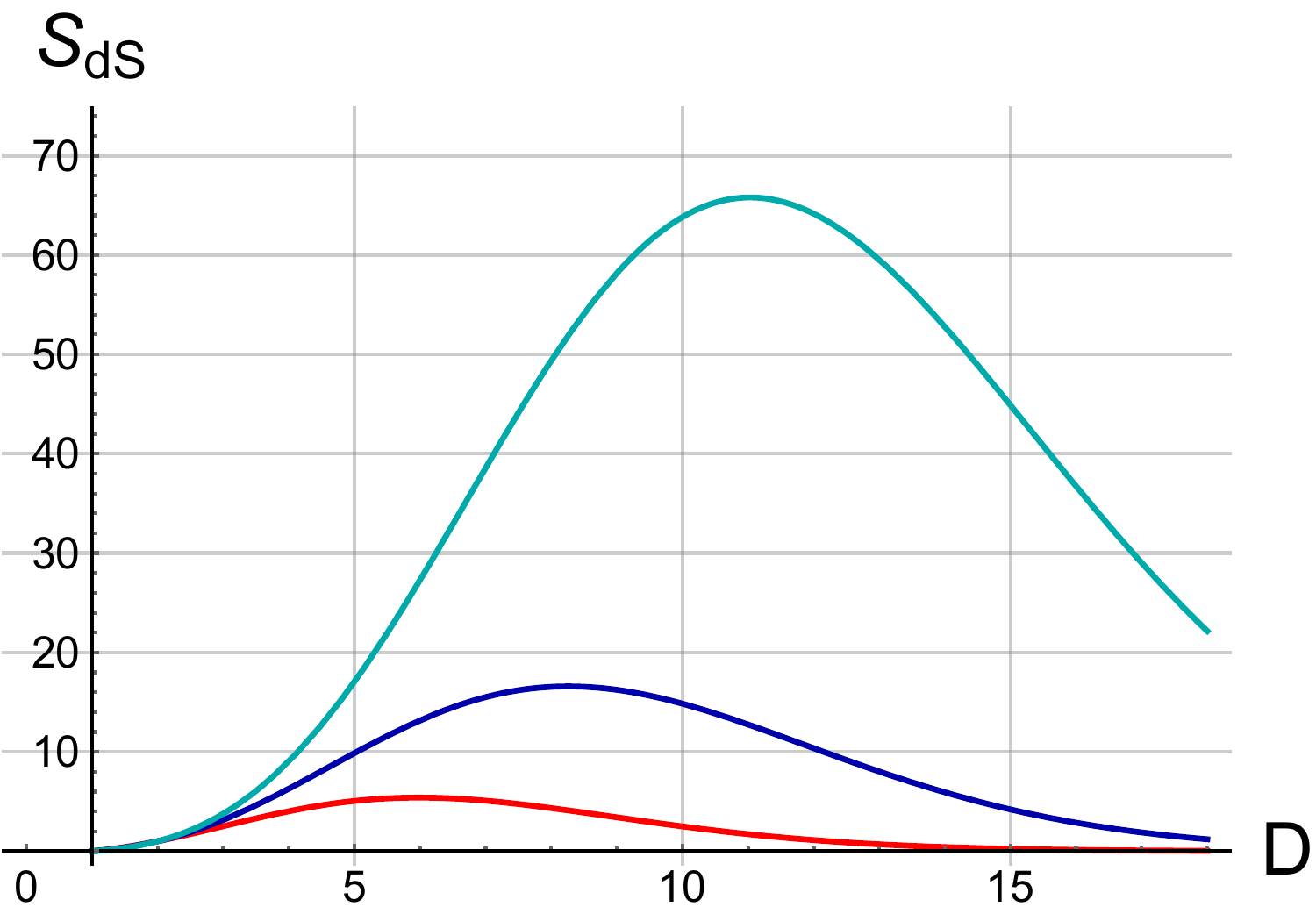} 
\qquad \includegraphics[width=70mm]
{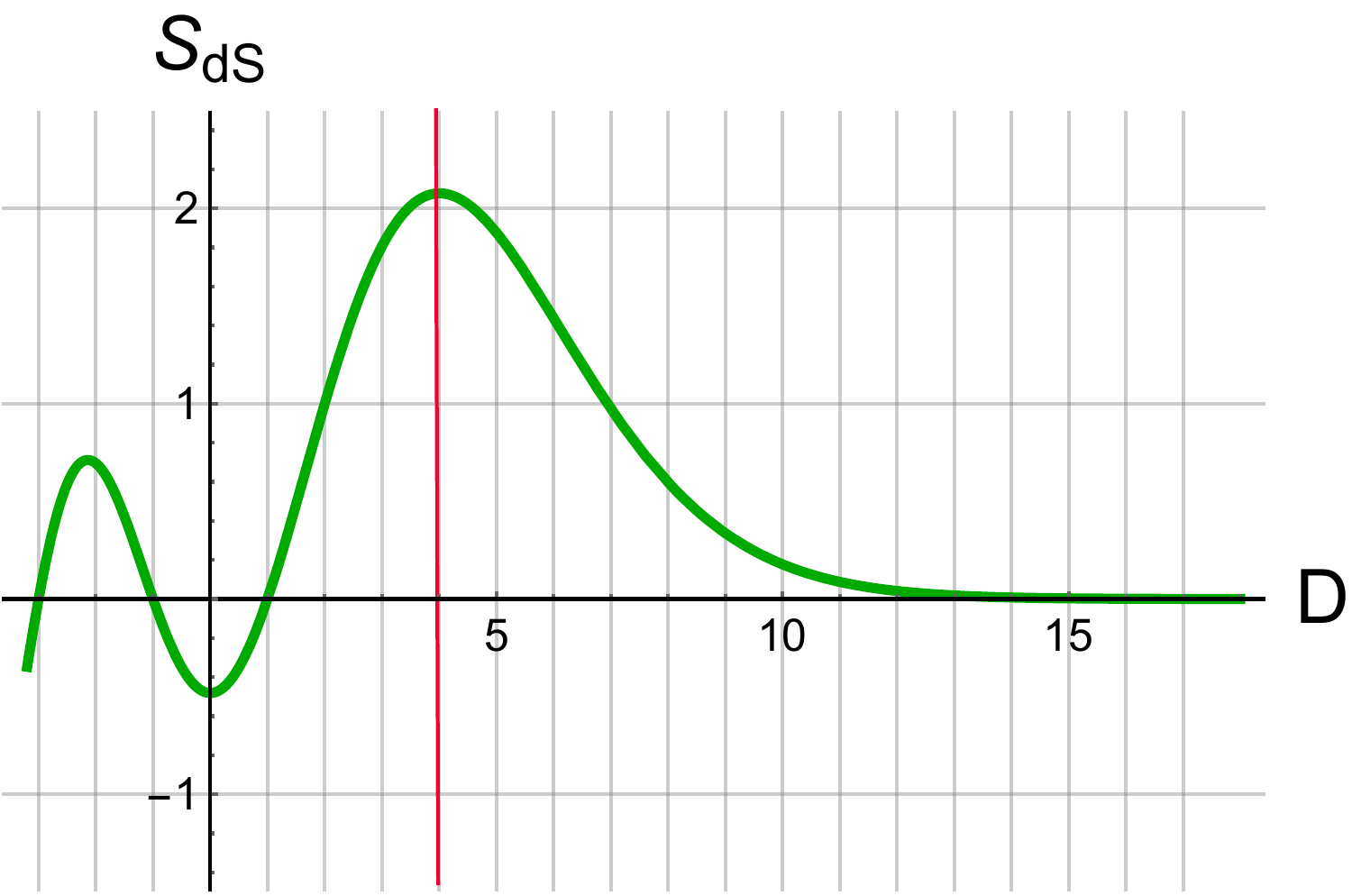} \\
A \hspace{210pt}B
\caption{A) The entropy $S_{dS}=S(\ell,D)$ as function of $D$ for different $\ell$: $\ell=1.2$ cyan line,
$\ell=1.0$ blue line and $\ell=0.8.$ red line. B)
$S_{dS}$ as function of $D$ for  $\ell=\ell_0=0.575$. The vertical red line shows the location of maximum entropy for $D=4$.}\label{fig:Sd}
 \end{figure}

For  extremum of the entropy with respect to
  $D$ we have
\be\frac{\partial S(\ell,D)}{\partial D}
=0.\ee
Solving this equation 
for $D=4$ with respect to $\ell_0$ we find 
\be
\label{ell0m}
\ell_0^2=\frac{1}{\pi} \exp \{\psi(3/2)\}\approx 0.3301,\quad  \mbox{i.e.}\quad \ell_0\approx 0.5745\ee
In Fig.\ref{fig:Sd}.A we plot the entropy as function of $D$ for different $\ell$. 
We see that $S(\ell,D)$ for fixed $\ell$ and varying $D$ in the region $D>1$ define  positive functions with one maximum, which location depends on $\ell$. \\

The blue curve in  Fig.\ref{fig:Sd}.A shows that for $\ell=1.0$ the maximum of the entropy is at $D=8.257$, since $S(1,D)=\Omega _{D-2}/4$ and when $D$ varies, the maximum area of unit spheres
$S^{D}$ is reached at $D=6.257$, see Sect.\ref{CND-va} and Fig.\ref{fig:Omega}.\\

As we see from above formula, one can take $\ell =\ell_0$ such that for this value of $\ell_0$
the maximum in $D$ is realized at $D=4$, see Fig.\ref{fig:Sd}.B.
 The corresponding cosmological constant is 
\be\label{Lambda0m}
\Lambda= \frac{3}{\ell_0^2}=3\pi\,\exp\{-\psi(3/2)\}\approx 9.087,
\ee
here we mean the Planck units, so $\Lambda\approx 9.087 M_{pl}^2=4.3 \cdot 10^{-15} \cdot $kg$^2$.

\begin{figure}[h]
    \centering
    \includegraphics[width=70mm]
{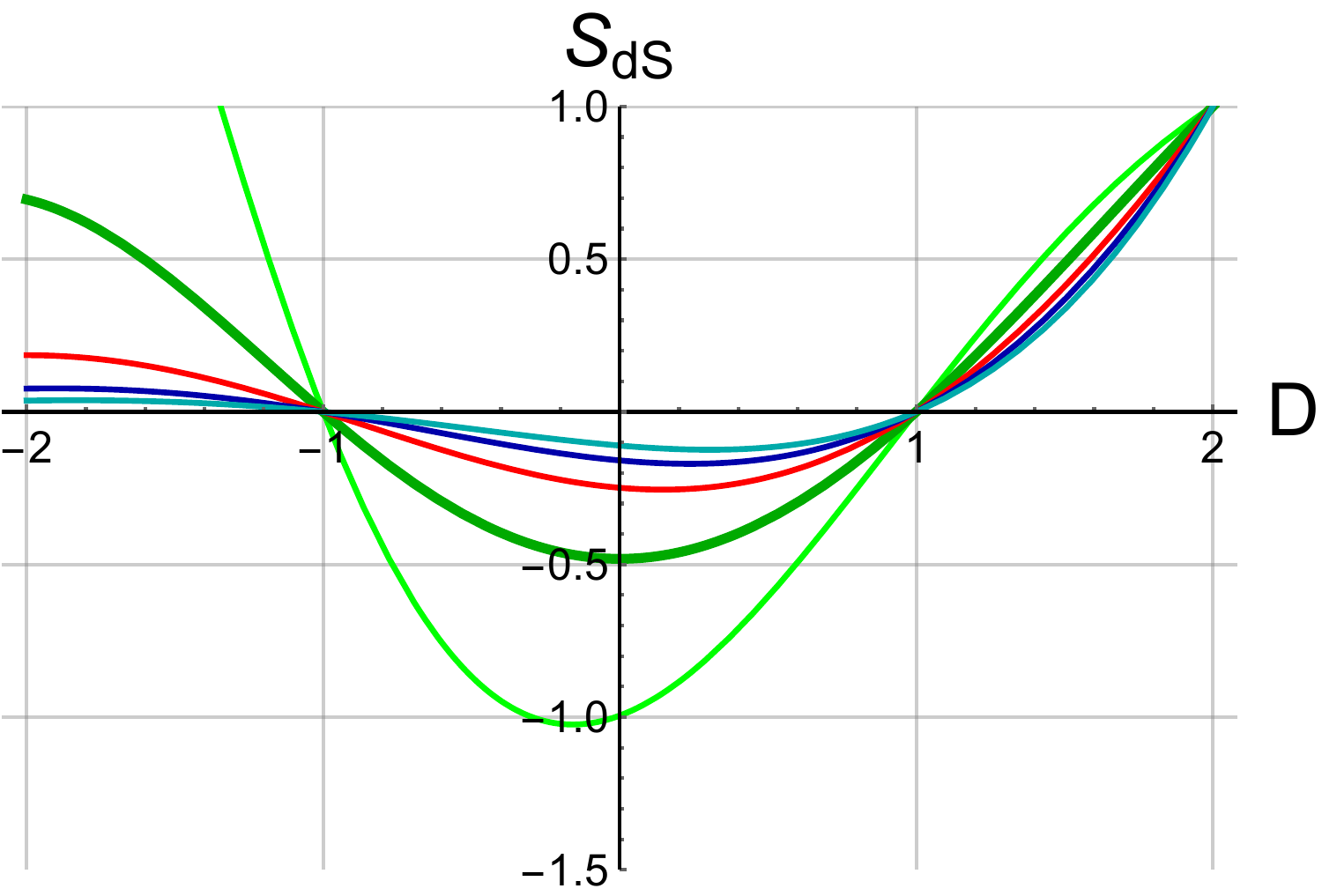} 
\qquad \includegraphics[width=70mm]
{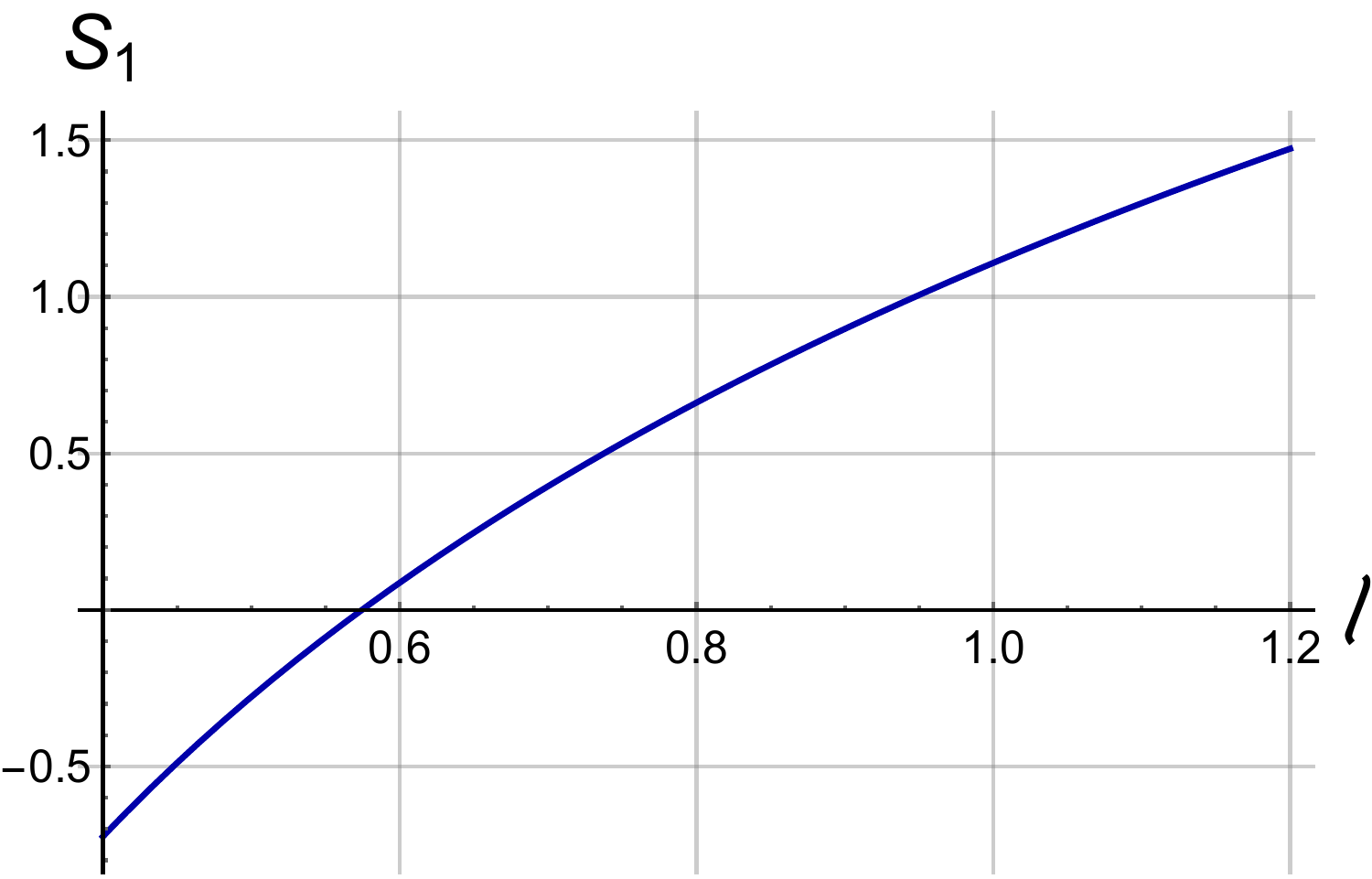} \\
A \hspace{210pt}B
\caption{A)  The entropy $S_{dS}=S(\ell,D)$ as function of $D$ for different $\ell$ near $D=0$: $\ell=1.2$ the cyan line,
$\ell=1.0$ the blue line, $\ell=0.8.$ the red line, $\ell=\ell_0$ darker green line and $\ell=0.4$ the green line. B) $S_1= \log (l^2\pi)-\psi
   \left(-\frac{1}{2}\right)$ as function of  $\ell$.  We see that the line crosses the horizontal axis at the point $\ell=\ell_0=0.575$.
   }\label{fig:S1D}
 \end{figure}

For $D=1$ and all $\ell$ the entropy is equal to zero and for $0<D<1$ it becomes negative for all $\ell$, see Fig.\ref{fig:S1D}.A.

It is interesting to note that the entropy as function of $D$ and fixed $\ell$ given by equation \eqref{ell0}
has also a minimum exactly at the zero dimension, $D=0$. Indeed, 
\be
[\log (\pi \ell^2)-\psi
   \left(\frac{x-1}{2}\right)]\Big|_{\ell=\ell_0,\, x=0}=0,\ee
 see Fig.\ref{fig:S1D}.B.  The value of entropy for $\ell=\ell_0$ and $D=0$
   is
   negative
  \be
  S(\ell_0,0)=-0.482084.\ee 
For other $\ell$, the entropy values for $D=0$ are also negative, but the minimum  values of entropy
are implemented for other values of $D$, see Fig.\ref{fig:S1D}.A.

\newpage
\section{Conclusion}

 In this note the inflationary cosmological constant $\Lambda$ of early universe is computed. It is given by eq.\eqref{Lambda0m}. For this computation the principal of maximum entropy and analytical continuation on spacetime dimension are used. Note that analytical continuation on spacetime dimension
 is widely used  in quantum field theory \cite{tHooft:1973wag} and  statistical mechanics \cite{Wilson:1973jj}.\\

The problem of initial conditions for inflationary models \cite{Starobinsky80,Guth81,Linde82}
is considered in \cite{Linde:2017pwt,Brandenberger:2016uzh}.
One can  use this cosmological constant as  the initial conditions for inflationary models. In particular, for  the Friedman equation for  the Hubble parameter we mean 
 the following initial condition:
\be
H(0)=\frac1{\ell_0},\ee
where $\ell_0$ is given by equation \eqref{ell0m}.
\\

As a final remark let us note that we have been discussing the creation of the universe from ``nothing" after the Planck era. For a consideration of quantum cosmology in the Planck era based on p-adic number, see \cite{Arefeva:1990cy,Dragovich:2022,Dragovich:2023} and references therein. \\

    \section*{Acknowledgements} I would like to thank I. Aref'eva, V. Belokurov,  V. Berezin, B. Dragovich,  E. Dynin, V. Frolov,  R. Kerner,
   V. Zagrebnov for fruitful discussions. This work is supported by  the Russian Science Foundation (project
19-11-00320, V.A. Steklov Mathematical Institute)

\appendix
\section{Appendix.}\label{CND}

\subsection{Area  of sphere and negative dimensions }\label{CND-va}
\begin{figure}[h]
    \centering
    \includegraphics[width=80mm]
{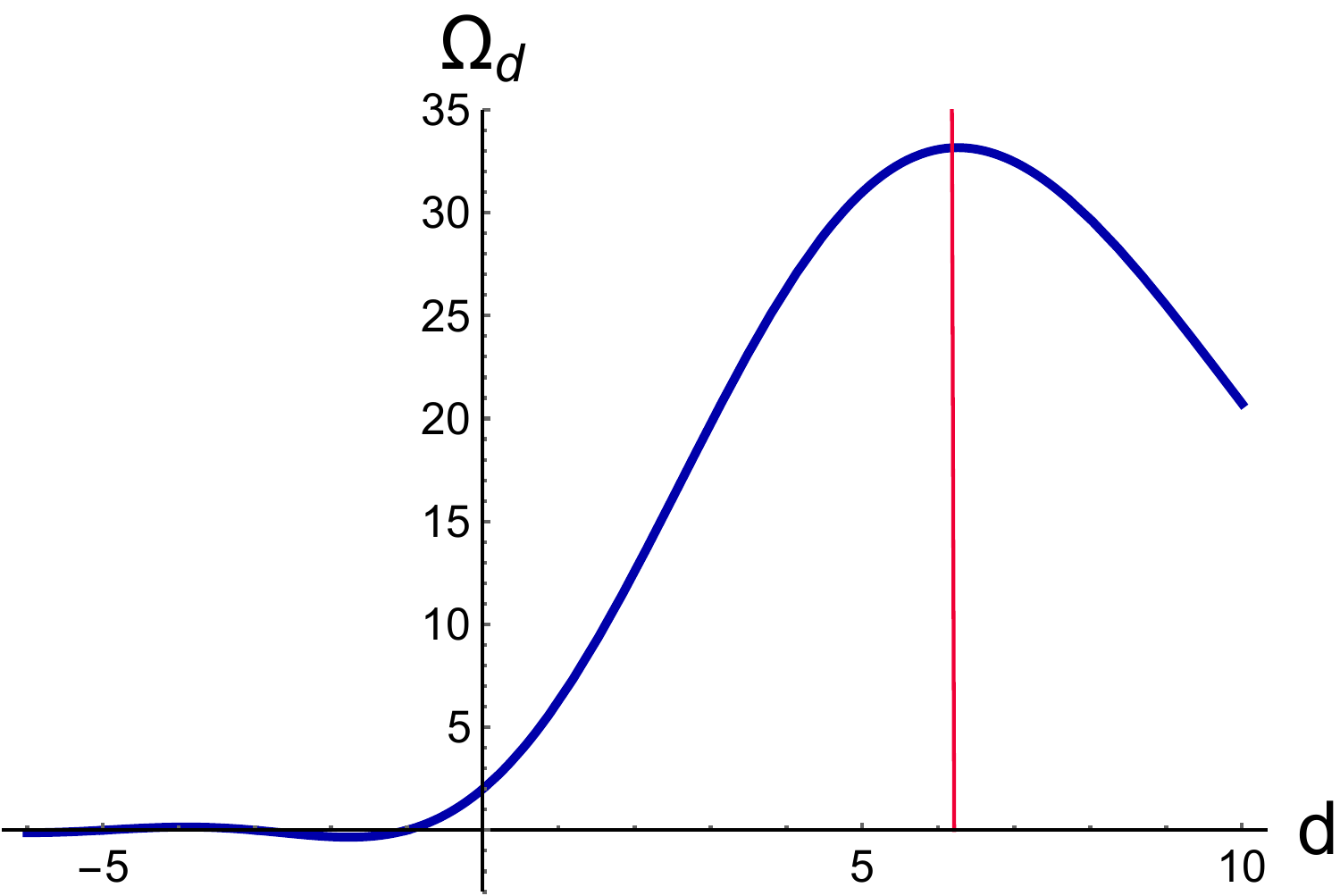}\qquad\includegraphics[width=60mm]
{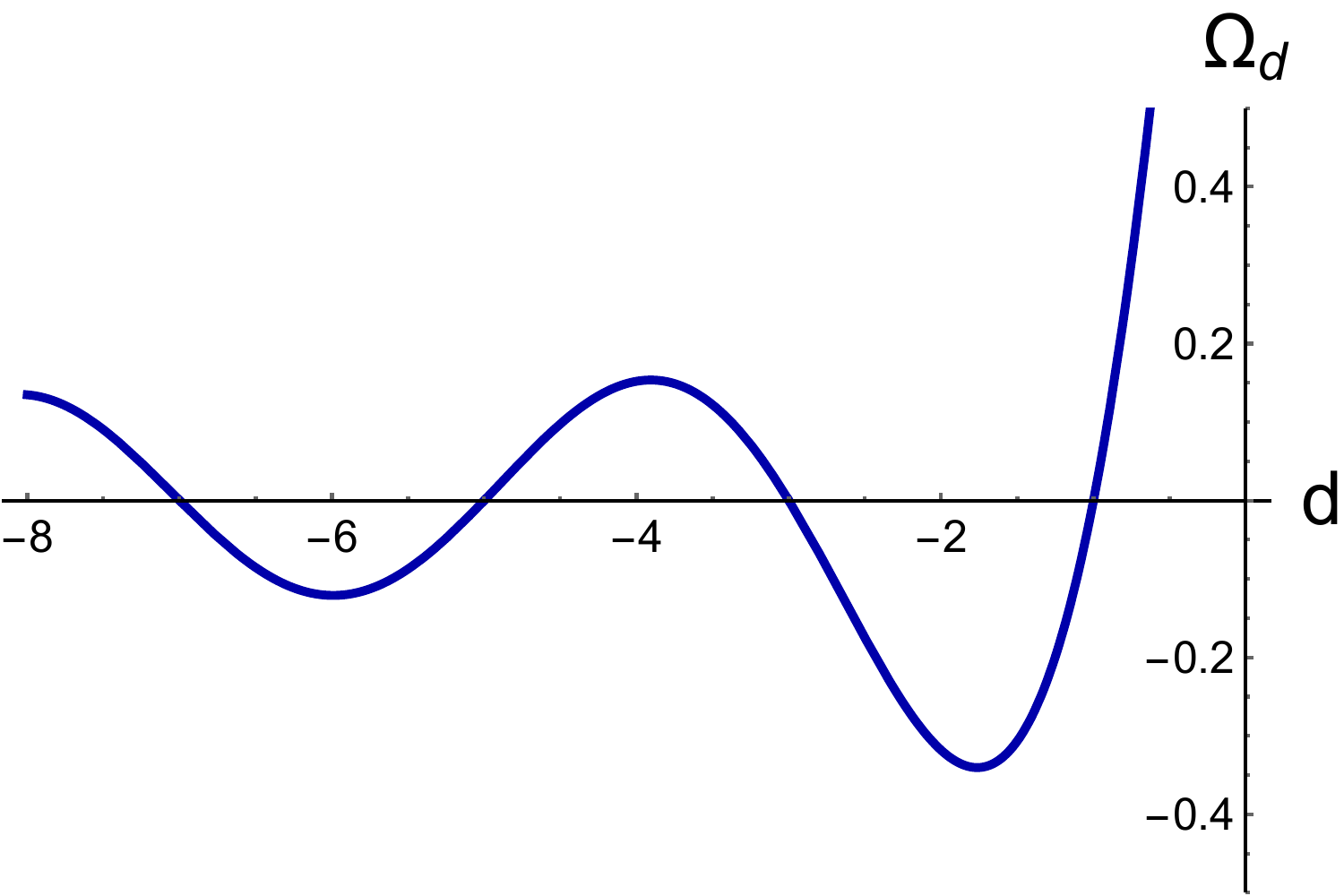}  \\
{\bf A}\hspace{150pt}{\bf B}
\caption{A) The area   $\Omega_d$ of $d$ dimensional  sphere $S^{d}$ as function of $d$.
B) Zoom of A) for negative $d$.
}\label{fig:Omega}
 \end{figure}

 The area of the unit d-dimensional sphere $S^d$ is
\be
     \Omega _d=\frac{2 \pi ^{(d+1)/2}}{\Gamma(\frac{d+1}{2})}, 
     \ee
     where $\Gamma(z)$ is the gamma function. The functions $L^d$ and  $\Omega _d$ admit  analytical continuations for all complex values of $d$, in particular to negative $d$. In the Fig.\ref{fig:Omega}
     the form of the function $\Omega _d$ is shown for real $d$.
     
For  extremum of $\Omega _d$ to respect of
  $d$ we have
\be\frac{\partial \Omega _d}{\partial d}
=0,\ee
or explicitly,
\be
\frac{\pi ^{\frac{d+1}{2}} \left(\log (\pi )-\psi
   \left(\frac{d+1}{2}\right)\right)}{\Gamma
   \left(\frac{d+1}{2}\right)}=0,\ee
where $\psi (z)$ is the digamma function,
 \be\label{psi}\psi (z)
   = \frac {d}{d z}\ln \Gamma (z).
   \ee
A solution $d=d_0$ of  this equation for positive $d>0$
satisfies
\be\label{sol}
\log (\pi )-\psi
   \left(\frac{d_0+1}{2}\right)=0\ee
   and $d_0=6.257$. At this point the $\Omega_d$ reaches the local maximum, see Fig.\ref{fig:Omega}.A.\\
   
\begin{figure}[h]
    \centering
    \includegraphics[width=100mm]
{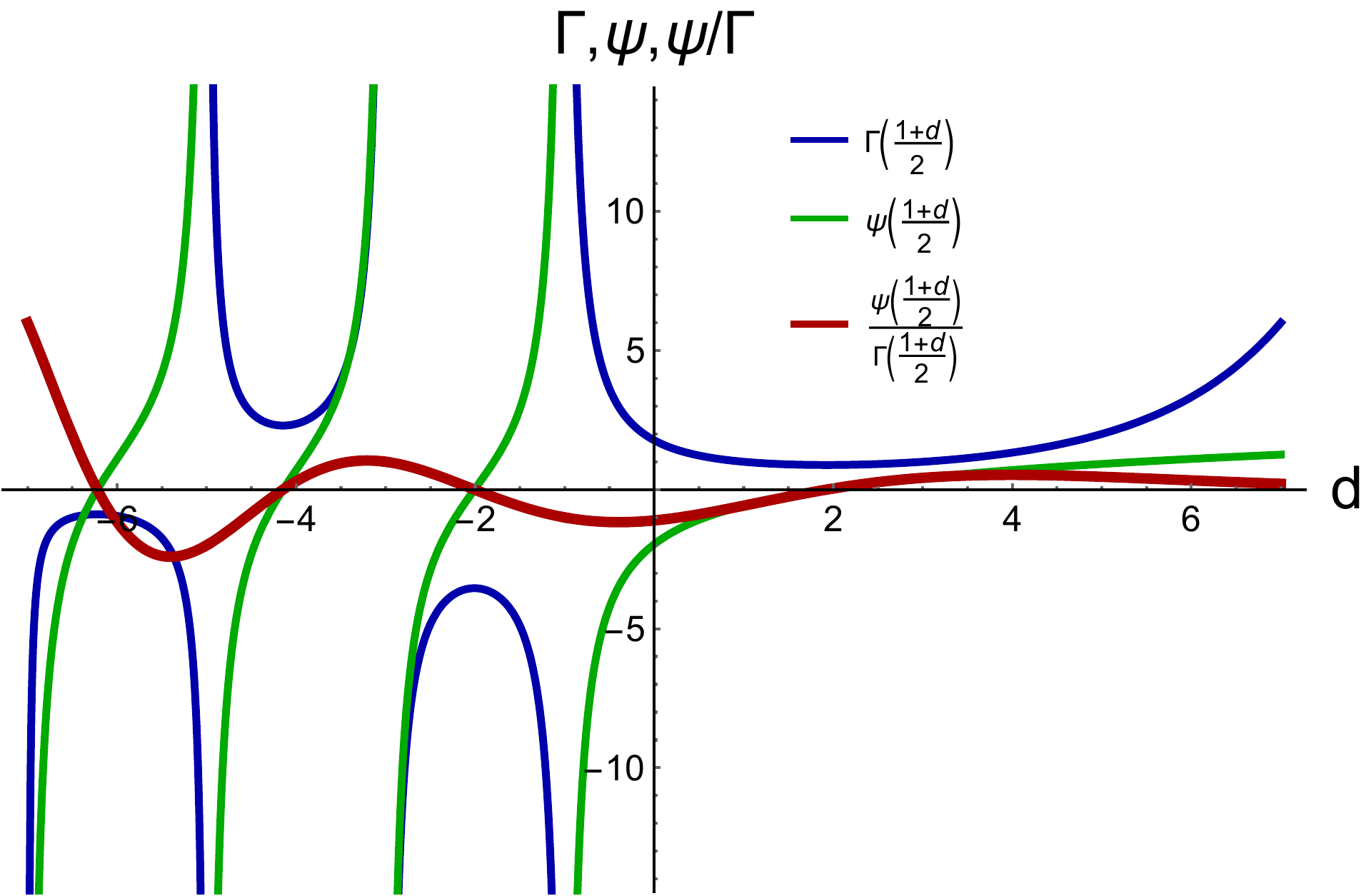}  
\caption{$\Gamma(\frac{d+1}{2})$, $\psi(\frac{d+1}{2})$ and $\psi(\frac{d+1}{2})/\Gamma(\frac{d+1}{2})$ as function of $d$.
}\label{fig:Gamma-Psi}
 \end{figure}

   There are other solutions of \eqref{sol} located at  negative $d$, $d=-1.77, -3.93, -6.01,...$, see Fig.\ref{fig:Omega}.B. Note that  negative odd $d$, $d=-2n-1$, do not provide the extremum of $\Omega_n$ since at these points  infinities of $\Gamma$ are compensated by infinities of $\psi$, see Fig.\ref{fig:Gamma-Psi}.
 
\end{document}